\documentclass[%
 aip,
 amsmath,amssymb,
 reprint,%
]{revtex4-1}

\usepackage{graphicx}
\usepackage{dcolumn}
\usepackage{bm}

\usepackage[utf8]{inputenc}
\usepackage[T1]{fontenc}
\usepackage{mathptmx}
\usepackage{etoolbox}
\usepackage{adjustbox}
\usepackage{hyperref}
\usepackage{cleveref}
\DeclareUnicodeCharacter{2212}{-}

\makeatletter
\def\@email#1#2{%
 \endgroup
 \patchcmd{\titleblock@produce}
  {\frontmatter@RRAPformat}
  {\frontmatter@RRAPformat{\produce@RRAP{*#1\href{mailto:#2}{#2}}}\frontmatter@RRAPformat}
  {}{}
}%
\makeatother
\begin{document}

\preprint{AIP/123-QED}

\title[Dipole and Cluster Dipole States of HPGe detectors at Cryogenic temperature]{Development of low-threshold detectors for low-mass dark matter searches with a
p-type germanium detector operated at cryogenic temperature}
\author{Mathbar Singh Raut}
 \affiliation{University of South Dakota,Vermillion, SD, 57069}
\author{Dongming Mei}%
 \email{dongming.mei@usd.edu}
$\affiliation{ 
University of South Dakota}$
\author{Sanjay Bhattarai}
$\affiliation{University of South Dakota}$
\author{Rajendra Panth}
$\affiliation{University of South Dakota}$
\author{Kyler Kooi}
$\affiliation{University of South Dakota}$
\author{Hao Mei}
$\affiliation{University of South Dakota}$
\author{Guojian Wang}
$\affiliation{University of South Dakota}$
\date{\today}

\begin{abstract}
 This study investigates new technology for enhancing the sensitivity of low-mass dark matter detection by analyzing charge transport in a p-type germanium detector at 5.2 K. To achieve low-threshold detectors, precise calculations of the binding energies of dipole and cluster dipole states, as well as the cross-sections of trapping affected by the electric field, are essential. The detector was operated in two modes: depleted at 77 K before cooling to 5.2 K and cooled directly to 5.2 K with various bias voltages. Our results indicate that the second mode produces lower binding energies and suggests different charge states under varying operating modes. Notably, our measurements of the dipole and cluster dipole state binding energies at zero fields were $8.716\pm 0.435$ meV and $6.138\pm 0.308$ meV, respectively. These findings have strong implications for the development of low-threshold detectors for detecting low-mass dark matter in the future.

\end{abstract}

\maketitle

\section{Introduction}\label{sec1}

The interaction between dark matter (DM) and ordinary matter results in only a small amount of energy being deposited through nuclear or electron recoil, which is limited to weak elastic scattering processes~\cite{article, Armengaud2018}. Therefore, detectors with extremely low energy thresholds are required to detect DM~\cite{Zhao2013, Wei2018, Essig2012}. Despite the modest mass of MeV-scale DM, its recent recognition as a potential DM candidate has generated interest. Unfortunately, current large-scale experiments are unable to detect MeV-scale DM due to its low mass. To detect MeV-scale DM, new detectors with sub-eV thresholds are needed as both electronic and nuclear recoils from MeV-scale DM range from sub-eV to $100$ eV~\cite{PhysRevD.105.042006, Petricca2020, Sundqvist2009}.

The detection of low-mass DM using conventional techniques is challenging. However, germanium (Ge) detectors offer a promising solution as they have the lowest energy threshold among current detector technologies, making them ideal for low-mass DM searches \cite{Armengaud2018, Agnese2019, Richard2003, Severijns2017}. Ge has a band gap of $0.7$ eV at $77$ K, and an average energy of $3$ eV is required to generate an electron-hole pair \cite{Wei2017}. This lower band gap in Ge is very favourable for the detection of low-mass DM. Furthermore, proper doping of the Ge detector with impurities can expand the parameter space for low-mass DM searches. Shallow-level impurities in Ge detectors have binding energies of about $0.01$ eV, which can form dipole states and cluster dipole states at temperatures below $10$ K \cite{Mei2022, 10.1088/1361-6471/acc751}. These states have even lower binding energies than the impurities themselves, offering a potential avenue for detecting low-mass DM. Although the binding energies of impurities in Ge are well understood \cite{Zakoucky2004, Naigh1999}, the binding energies of the dipole states and cluster dipole states near helium temperature is still poorly understood.

\begin{figure}[h]%
\centering
\includegraphics[width=0.45\textwidth]{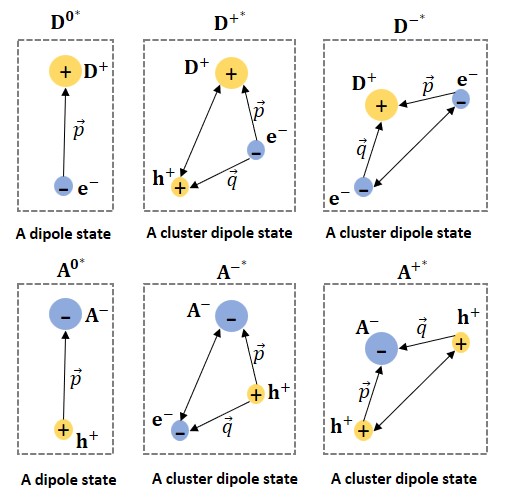}
\caption{An illustration of the processes that lead to the development of excited dipole states and cluster dipole states
in an n-type (upper) and a p-type (lower) Ge detector operated
at temperatures below 10 K, where $\vec{p}$ and $\vec{q}$ are the
corresponding dipole moments~\cite{Mei2022}.}
\label{fig1}
\end{figure}

As temperatures approach liquid helium levels, any remaining impurities in Ge detectors freeze out of the conduction or valence band and transit into localized states, forming electric dipoles ($D^{0*}$ for donors and $A^{0*}$ for acceptors) or neutral states ($D^0$ and $A^0$) \cite{Mei2022}. These dipole states have the ability to trap charge carriers and can form cluster dipole states $(D^{+}$ and $D^{-}$ for donors, and $A^{+}$ and $A^{-}$ for acceptors) \cite{Mei2022}. Figure~\ref{fig1} depicts the formation of dipole states and cluster dipole states at temperatures below 10 K. 

Previous studies by Mei et al have thoroughly examined this phenomenon \cite{Mei2022}. When an alpha particle ($\alpha$) from $^{241}Am$ decay is directed towards a Ge detector, it creates electron-hole pairs within a range of $10$ $\mu$m from the detector's surface \cite{Ziegler2010, Arnquist2022, Leon2012}. By operating the detector at a cryogenic temperature of approximately $4$ K and applying a positive or negative bias voltage to the bottom of the detector, only one type of charge carriers is drifted through it. To study the binding energy of the formed dipole states and cluster dipole states, these drifted charge carriers undergo a dynamic process of elastic scattering, trapping, and de-trapping. In this experiment, a p-type Ge detector is run in two different modes with different bias voltages while being cooled to cryogenic temperature. Similar experiment and its results for an n-type Ge detector operating in these two modes have already been published \cite{Bhattarai2023}.

\section{Experimental Procedure}\label{sec2}
The state-of-the-art infrastructure at USD for crystal growth and detector development includes a zone refining process for highly purifying commercial ingots, which can be used for crystal growth with the Czochralski method \cite{Raut2020, Yang2014, Wang2012, Bhattarai2020}. This enables the USD detector fabrication lab to produce superior homegrown crystals that are utilized for creating p-type (RL) detector with a net impurity concentration of $6.2\times10^9/cm^3$ and dimensions of $18.8$ mm $\times17.9$ mm$\times10.7$ mm.
The detector was fabricated using a sputtering technique, which deposits an amorphous Ge layer on the top, sides, and bottom of the detector to form electrical contacts. The detailed fabrication process has been published in our previous work titled "Fabrication and Characterization of High-Purity Germanium Detectors with Amorphous Germanium Contacts" by X.-H. Meng et al~\cite{meng}.
To ensure optimum electrical performance, an amorphous Ge passivation layer of $600$ nm is applied to the surface of the Ge crystal as the electrical interface to successfully block surface charges \cite{Wei2017, Panth2021}.

The detector is mounted inside a pulse tube refrigerator, which cools it down to nearly liquid helium temperature from room temperature. To ensure accurate temperature measurements, we have installed two temperature sensors inside the detector housing. One sensor is placed at the bottom of a copper plate on which the detector rests on a thin indium foil, while the other is located on top of another copper plate close to the top surface of the detector. By positioning the detector between these two sensors, we can measure its temperature with an accuracy of 0.5 K. The temperature readings from the two sensors are always within 0.5 K of each other.

We chose a working temperature of 5.2 K based on the capacitance measurements presented in Ref.\cite{Mei2022}. The capacitance measurements indicate that the capacitance remains constant when the temperature is below 6.5 K. To ensure that the capacitance remains stable and to err on the side of caution, we chose a working temperature of 5.2 K.

In the experiment, an alpha source ($^{241}$Am) was placed near the detector inside a cryostat to measure the energy deposition of $\alpha$-particles, creating localized electron-hole pairs near the top surface of the detector. By applying a negative bias voltage to the bottom of the detector, the holes are drifted through the detector. The experiment was conducted using two modes of operation.

In Mode 1, the RL detector was operated at 77 K with a depletion voltage of -400 V and an operational voltage of -1200 V. An $\alpha$-source ($^{241}$Am) emitting $\alpha$-particles with energy of $\sim5.5$ MeV was placed in close proximity to the detector inside the cryostat. The resulting energy spectrum was measured to detect the energy deposition of the $\sim5.5$ MeV $\alpha$-particles, which was visible as a $3.92$ MeV energy peak due to energy loss en route to the detector's active region. The negligible detector charge trapping at 77 K with a bias of -1200 volts made the $3.92$ MeV energy deposition an ideal reference for determining the energy deposition of $\sim5.5$ MeV alpha particles in the p-type detector without charge trapping. To calculate the charge collection efficiency, the measured alpha energy peak was divided by $3.92$ MeV for a specific bias voltage.

 The detector was cooled down to 5.2 K and maintained fully depleted by a negative bias voltage of -1200 V. Following $\alpha$ energy deposition on the detector's surface, the resulting holes began to drift across the detector. At this temperature, space charge could trap holes, leading to the formation of electric dipole states. To investigate these states, the detector was subjected to decreasing bias voltages ranging from -1200 V to -200 V. Energy deposition histograms of alpha particles were recorded every 2-3 minutes for a duration of 60 minutes at each bias voltage, enabling the collection of data on the dipole states and their properties.

When operating in Mode 2, the detector was grounded during the cool-down process and the detector was cooled immediately to 5.2 K without any bias voltage applied. Once the temperature reached 5.2 K, a negative bias voltage was gradually applied from the bottom of the detector, creating an electric field that caused the surface-generated holes to drift across the detector. The energy spectrum measurements were taken using bias voltages of -200 V, -300 V, -600 V, -900 V, and -1100 V. As in Mode 1, data was collected for 60 minutes at each bias voltage, with histograms of energy deposition by alpha particles recorded every 2-3 minutes. 

\begin{figure}[h]%
\centering
\includegraphics[width=0.45\textwidth]{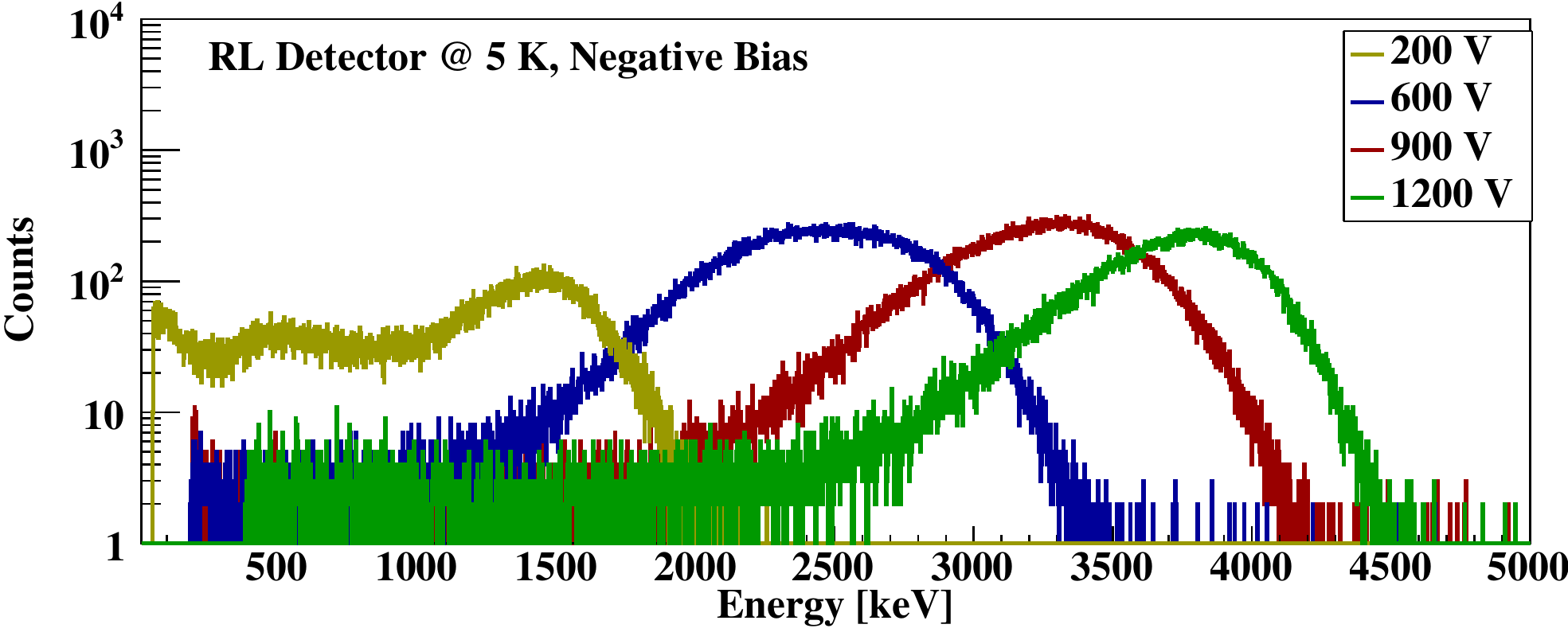}
\caption{The energy deposition of $\sim5.5$ MeV $\alpha$ particles in an
p-type detector operating in Mode 1.}
\label{fig2}
\end{figure}

\begin{figure}[h]%
\centering
\includegraphics[width=0.45\textwidth]{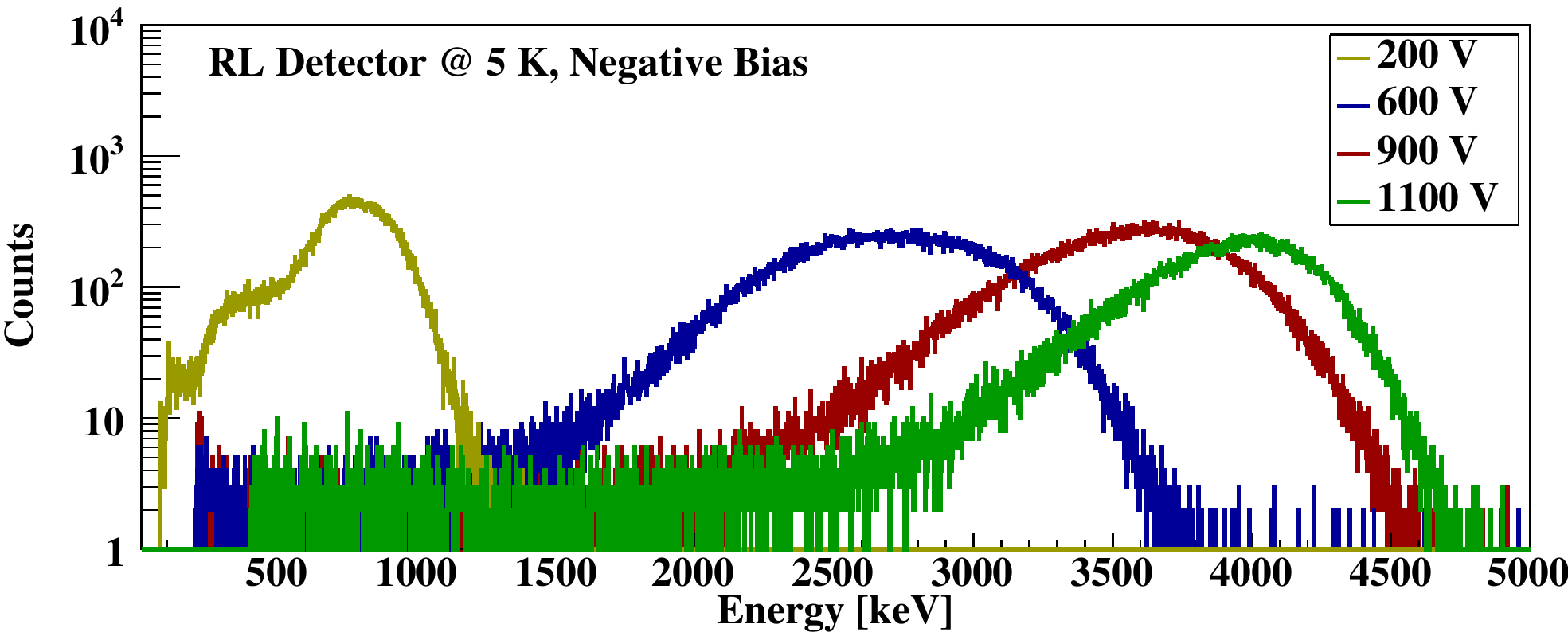}
\caption{The energy deposition of $\sim5.5$ MeV $\alpha$ particles in an
p-type detector operating in Mode 2.}
\label{fig3}
\end{figure}

Figures~\ref{fig2} and \ref{fig3} illustrate the energy deposition of the 5.5 MeV $\alpha$ particles emitted from $^{241}$Am decays when the detector was operated in Mode 1 and Mode 2, respectively. Both modes were designed to investigate different physical processes, which are explained below in terms of their physical processes.

\subsection{Mode 1}\label{subsec1}

A p-type planar detector is first cooled to $77$ K in this mode, and then a negative bias voltage is applied to the bottom of the detector, gradually raising it until the detector is completely depleted at -400 V. After that, the bias is elevated by an additional 800 volts to reach the operational voltage. The detector is then brought under the operational voltage while being cooled down to $5.2$ K. At $77
$ K, the depletion process causes all the free-charge carriers
to be swept away, leaving only the space charge states,
$A^-$, behind. Upon cooling to $5.2$ K, a trapping process occurs.
As the holes continue to drift across the detector, the de-trapping process occurs ~\cite{Mei2022}.
The key trapping and de-trapping processes are described below:
\begin{equation}
{h^+} + {A^-} \rightarrow A^0{^*};
{h^+} + {A^0{^*}} \rightarrow A^{-}+2{h^+}.
\label{eq1}
\end{equation}

When the detector is operated in negative bias at the bottom, the Coulomb force between the space charge states and the drifting holes occurs. The p-type planar detector's operation in this mode starts with the formation of dipole states through charge trapping. Charge de-trapping, also known as the second process, involves releasing trapped charge by ionizing the dipole states through impact ionization.

We can figure out the binding energy of the dipole states by looking at the time-dependent behavior of this de-trapping process.

\subsection{Mode 2}\label{subsec2}
The p-type planar Ge detector is directly cooled to $5.2$ K in this mode of operation, with no bias voltage provided. The detector is then biased to the required voltage level after cooling. Impurities in the Ge crystal freeze out of the conduction or valence band at very low temperatures, forming localized states that give rise to dipole states. As it is a p-type detector, the majority of these dipole states are $A^{0*}$~\cite{Mei2022}.
When an $\alpha$ source is positioned near the detector, electron-hole pairs are created on the detector's surface. The resulting holes then drift across the detector upon application of a negative bias voltage to its bottom. This initiates the following processes within the detector:
\begin{equation}
{h^+} + {A^{0^*}} \rightarrow{A^{+^*}};\\
{h^+} + {A^{+^*}} \rightarrow 2{h^+}+{A^{0^*}}.
\label{eq2}
\end{equation}

Operation of the detector in a negative bias mode leads to the production of cluster dipole states as the initial process. These states arise from the Coulomb forces exerted on the drifting holes, resulting in the trapping of charges. The second process involves the impact ionization of the cluster dipole states, leading to the de-trapping of charges. Charge production, generation, and transport occur dynamically within the detector, and the study of the time-dependent de-trapping of charges through the impact ionization of cluster dipole states can provide insights into their binding energies.

It is important to note that when comparing the two operational modes, Mode 2 creates the dipole states at $5.2$ K without the requirement for a bias voltage to be applied. When the holes cross the detector, these dipole states quickly trap the charges, resulting in a shorter trapping time and lower binding energy. In contrast, when an applied bias voltage causes holes to drift across the detector, Mode 1 produces the dipole states in the space charge area. The trapping time is therefore anticipated to be longer and the binding energy of the dipole states to be larger than that of the cluster dipoles.

\section{Physical Model}\label{sec3}
The following concepts highlight the physics model employed in this investigation. A cryogenically cooled HPGe detector placed near an $\alpha$ source causes free charge carriers to drift along the detector and become captured in electric dipole states, forming cluster dipole states. The increase in cluster dipole states is accompanied by a decrease in electric dipole states, indicating a reduction in charge trapping. Charge carriers trapped in the cluster dipole states begin to emit from the traps upon continuously biasing the detector. The emission rate of these charge carriers is time-dependent and reaches saturation once all of the trapped carriers have been released. The emission rate can be determined using references such as ~\cite{Lee1999, Agnese2014}:
\begin{equation}
{e_n} = {\sigma_{trap}v_{th}N_\nu\exp\left(\frac{-E_B}{k_BT}\right)},
\label{eq3}
\end{equation}
where $\sigma_{trap}$ is the trapping cross-section,
$v_{th}$ is the thermal
velocity, $N_{\nu}$ is the effective density of states of holes
in the valence band, $E_B$ is the binding energy of the trapped charge carriers, $k_B$ is the Boltzmann
constant, and $T$ is the temperature of the detector.

Equation~\ref{eq3} can be used to obtain the binding energy of dipole states or cluster dipole states if the trapping cross-section ($\sigma_{trap}$) is known, provided that experimental data is used to determine $e_n$ directly, along with the values of $v_{th}$, $N_{\nu}$, and $T$. However, determining the value of $\sigma_{trap}$ requires additional calculations, as will be explained below.

The relationship between the trapping cross-section of charge carriers and the trapping length $\lambda_{trap}$ is described by the following equation~\cite{Mei2020, He2001}:
\begin{equation}
{\lambda_{trap}} = \frac{1}    {\left(\frac{{N_A+N_D}\pm \mid {N_A-N_D}\mid}{2}\right)\times \left(\sigma_{trap} \times\frac{v_{tot}} {v_d}\right)},
\label{eq4}
\end{equation}

Where $N_A$ and $N_D$ represent the p-type and n-type impurities,
respectively. Note that the method we used to determine these densities is consistent with that used in our previous publication~\cite{Mei2020}. Our findings indicate that there is a factor of approximately 10 difference between N$_A$ and N$_D$, which allows us to utilize N$_A$ – N$_D$ as a suitable representation of N$_A$. $v_{tot}$ is the total velocity of the drift holes,
and $v_d$ is the drift velocity, which is electric field dependent
and is given by~\cite{10.1088/1361-6471/acc751},
\begin{equation}
{v_{d}} \approx \frac{\mu_{0}E} {\left(1+ \frac{\mu_{0}E}{v_{sat}}\right)},
\label{eq5}
\end{equation}
where $\mu_0$ is the mobility of the charge carrier when
the field is zero and is equal to 
\begin{equation}
{\mu_{0}} = \frac{\mu_{0}(H)}{r},
\label{eq6}
\end{equation}

and $\mu_0(H)$
is the Hall mobility. The IEEE standard values for $\mu_0(H)$
and $r$ are $36000cm^2/Vs$ for electrons and $42000cm^2/Vs$
for holes and $0.83$ for electrons and $1.03$ for holes respectively~\cite{Quay2000,Abrosimov2020}.
The saturation velocity, $v_{sat}$, can be calculated
according to an empirical formula below~\cite{Mei2020}.

\begin{equation}
{v_{sat}} =\frac{v_{sat}^{300}}
{(1-A_\nu+A_\nu{\left(\frac{T}{300}\right))}}.
\label{eq7}
\end{equation}

The saturation velocity at $300K$ ($V_{sat}^{300}$) for electrons and
holes are $7\times10^6cm/s$ and $6.3\times10^6cm/s$ respectively. The
values of $A_\nu$ for electrons and holes are $0.55$ and $0.61$
respectively~\cite{Ahmed_2011}. 

Moreover, charge collection efficiency
$(\epsilon_h)$ of a planar HPGe detector is related to $\lambda_{trap}$ by ~\cite{Mei2020, He2001}
\begin{equation}
{\epsilon_{h}} = {\frac{\lambda_{trap}}{L}}
{(1-\exp\left(\frac{-L}{\lambda_{trap}}\right))},
\label{eq8}
\end{equation}
where $L$ is the detector thickness. 
For the known value of the net impurity concentration, and the thickness of the detector,
Equation~\ref{eq4} allows us to determine the charge trapping cross-section $(\sigma_{trap})$ in a planar Ge detector by determining the charge collecting efficiency $(\epsilon_h)$.

We can determine $\lambda_{trap}$ from equation ~\ref{eq8} using the calculated values of $\epsilon_h$ and the known detector thickness $(L)$.
The charge carriers' combined total velocity, or $(v_{tot})$, is made up of their thermal velocity$(v_{th})$, and their saturation velocity$(v_{sat})$.  Therefore, the electric field-dependent trapping cross-section $(\sigma_{trap})$ can be calculated by combining the equations for $\lambda_{trap}$  and $v_{tot}$~\cite{Mei2020, 2016PhDT........78P} described above.

When operating a p-type detector (RL detector) in both Mode 1 and Mode 2, the emission rate $(e_n)$ of charge carriers from the traps can be measured. To calculate the emission rate, a specific bias voltage is applied to the detector, and the slope of the energy versus time plot is used. By combining the measured value of $(e_n)$ with equation~\ref{eq3}, we can determine the binding energies of dipole states and cluster dipole states in the p-type Ge detector at cryogenic temperatures.

\section{Results and Discussion}\label{sec4}

\begin{figure}[htp] 
\centering
\includegraphics[width=0.45\textwidth]{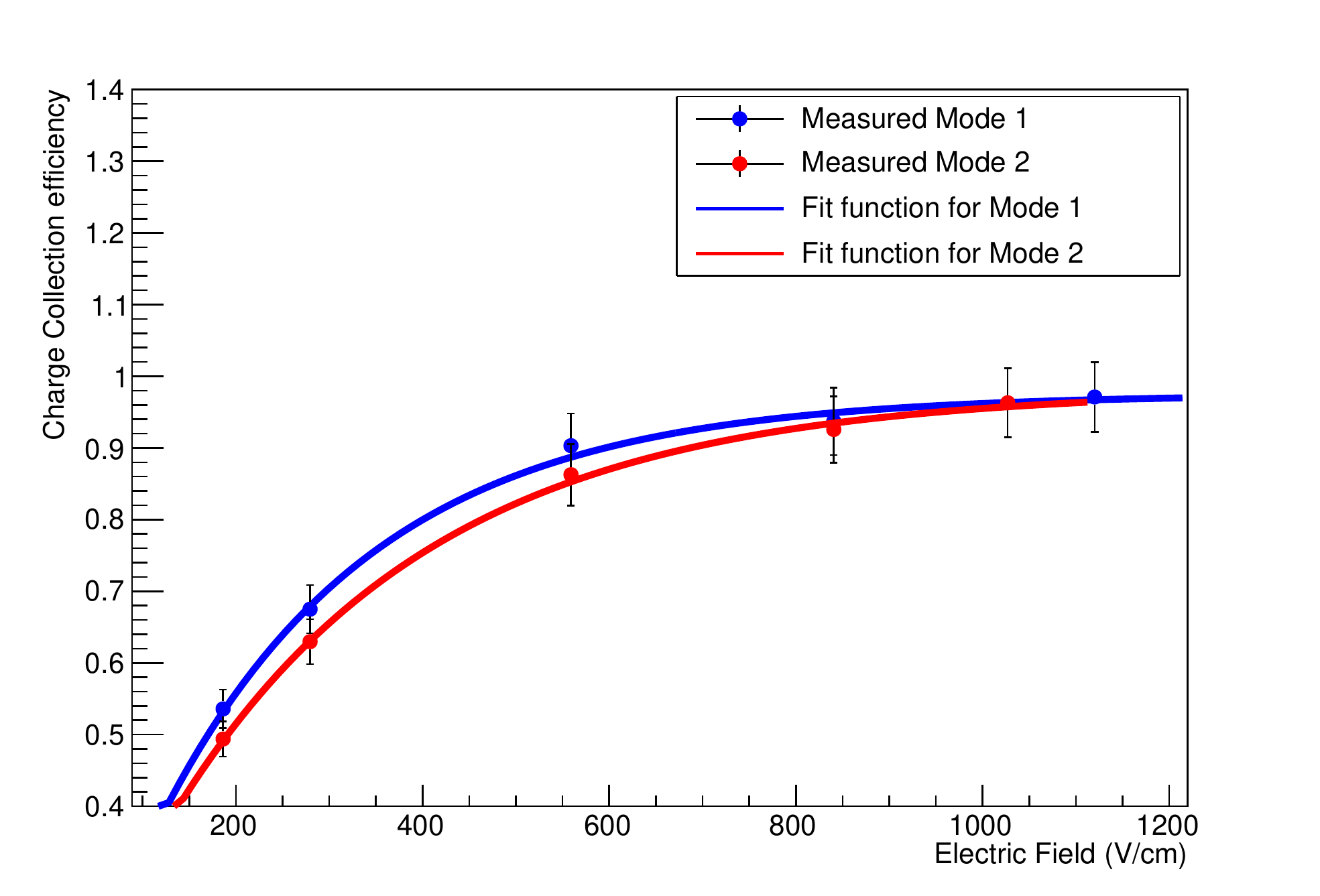}
\caption{The graph of charge collection efficiency ($\epsilon$) versus applied electric field ($E$) for detector RL has been plotted,
with errors taken into account. The error in $\epsilon$ is based on the measurement of the mean energy deposition, while the error in $E$ is largely influenced by the bias voltage applied. A fitting model, $\epsilon = p_0 +[(p_1 \times exp(−p_2 \times E)]$, was utilized to curve-fit the data, resulting in the following fitted parameters: For Mode 1: $p_0 =0.974\pm0.044$, $p_1 = -0.994\pm 0.0237$, and $p_2 = -0.00433\pm0.000145$. Similarly, for Mode 2:$p_0 =
0.9815\pm0.0586$,$p_1=-0.9553\pm 0.0164$,and $p_2=-0.0035\pm 0.00023$. }
\label{fig4}
\end{figure}

Figures~\ref{fig2} and~\ref{fig3} display the energy deposition from $\sim5.5$ MeV $\alpha$ particles in Mode 1 and Mode 2 of the RL detector, respectively. To determine the charge-collection efficiency of the detector, we compared the mean total energy deposited at 5.2 K with a certain bias voltage to the mean energy deposited at 77 K when the detector was depleted and operated with a bias voltage of -1200 volts. For example, the mean energy observed at 77 K with a bias voltage of -1200 V was $3.92$ MeV, whereas the mean energy observed at -200 V at 5.2 K was $2.09$ MeV, resulting in a charge collection efficiency of $53.3\% (\epsilon = $2.09$ MeV/$3.92$ MeV)$. We plotted the charge-collection efficiency as a function of the applied bias voltage in Figure~\ref{fig4}. 

\begin{figure}[htp] 
\centering
\includegraphics[width=0.45\textwidth]{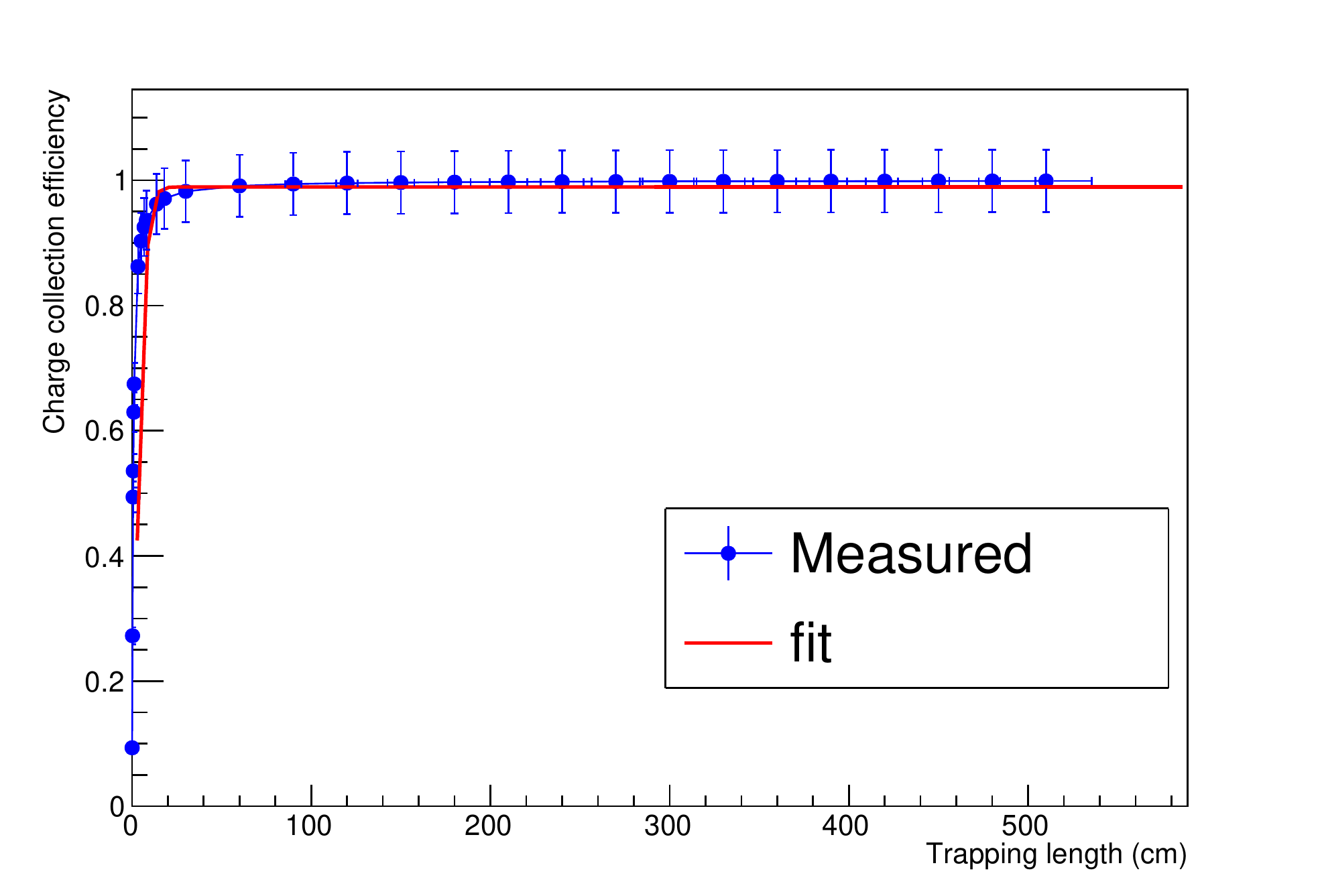}
\caption{The charge collection efficiency ($\epsilon$) of a p-type detector RL as a function of trapping length ($\lambda_{trap}$) has been plotted, taking into account the associated errors. The error in $\epsilon$ is determined from the measured mean energy deposition, while the error in $\lambda_{trap}$ is obtained through the propagation of error in the expression for $\lambda_{trap}$ given by ~\ref{eq8}. A fitting model of the form $\epsilon = \frac{p_0}{1+p_1\exp(-p_2\lambda_{trap})}$ has been applied to the data, resulting in the following fitted parameters for the p-type detector RL: $p_0=0.9893\pm0.049$, $p_1=4.764\pm0.238$, and $p_2=0.4335\pm0.0216$. This model provides a good fit to the experimental data and can be used to predict the charge collection efficiency for different values of $\lambda_{trap}$.}
\label{fig5}
\end{figure}

Using the thickness (L) of the detector (10.7 mm) and the charge-collection efficiencies obtained at various bias voltages, we calculated the trapping length ($\lambda_{trap}$) of the charge carriers with equation~\ref{eq4}. Figure~\ref{fig5} shows the charge collection efficiency versus the trapping length. 

\begin{figure}[htp] 
\centering
\includegraphics[width=0.45\textwidth]{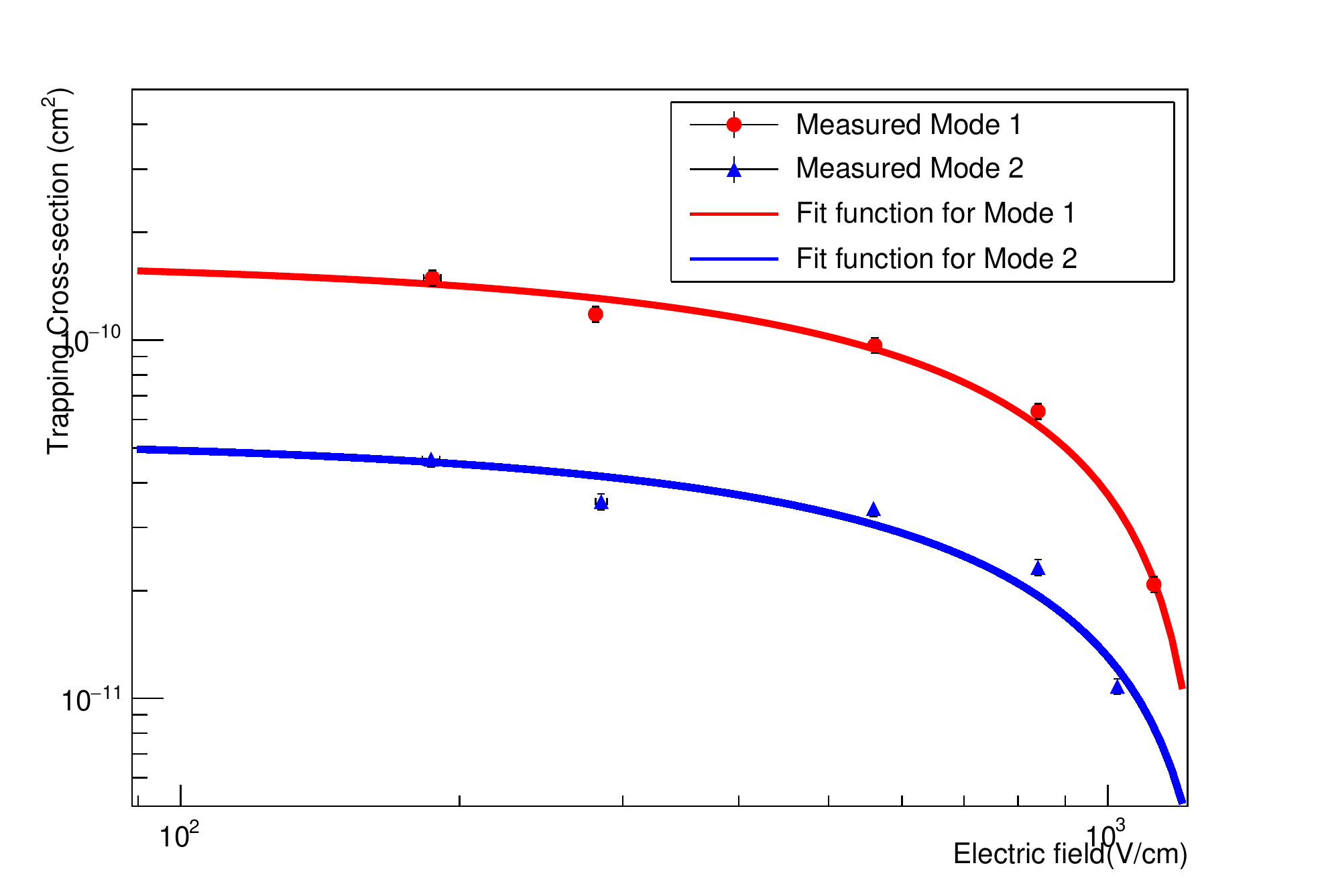}
\caption{The graph illustrates the relationship between the trapping cross-section ($\sigma_{trap}$) and the applied bias field ($E$) for detector RL in both Mode 1 and Mode 2, accounting for errors. The error in $\sigma_{trap}$ is determined using the propagation of error in equation~\ref{eq4}, while the error in $E$ primarily stems from the applied bias voltage. A fitting model of the form $\sigma_{trap} = p0 -[(p1)\times \exp(p2\times E)]$ was used to fit the data. For Mode 1, the following fitted parameters were obtained: $p0 = -1.038\times 10^{-8}\pm3.344 \times 10^{-10}$, $p1 = -1.055\times10^{-8}\pm4.336\times10^{-10}$, and $p2 = 1.245\times 10^{-5}\pm3.937\times 10^{-7}$. Similarly, for Mode 2, the fitted parameters were found to be $p0 = -8.596\times10^{-10} \pm 1.24\times10^{-11}$, $p1 = -9.124\times10^{-10}\pm8.645\times10^{-12}$, and $p2 = 4.527\times 10^{-5}\pm 1.726\times 10^{-7}$.}
\label{fig6}
\end{figure}
We measured the net impurity concentration of the detector to be $6.2\times10^9/cm^3$ and operated it at a temperature of $5.2$ K using the two modes described earlier. These values, along with other parameters presented in equations~\ref{eq5},\ref{eq7},\ref{eq8}, were used to calculate the trapping cross-section of the trap centers. The relationship between the trapping cross-section and the applied electric field is illustrated in Figure~\ref{fig6}. It's worth noting that these trapping cross-sections should be considered as effective trapping cross-sections, as there is no known way to separate the various processes of trapping. When comparing the effective trapping cross-section of the trap centers to the charge states in V. N. Abakumov et al.~\cite{abak}, we found that the results are in the same order of magnitude.

\begin{figure}[htp] 
\centering
\includegraphics[width=0.45\textwidth]{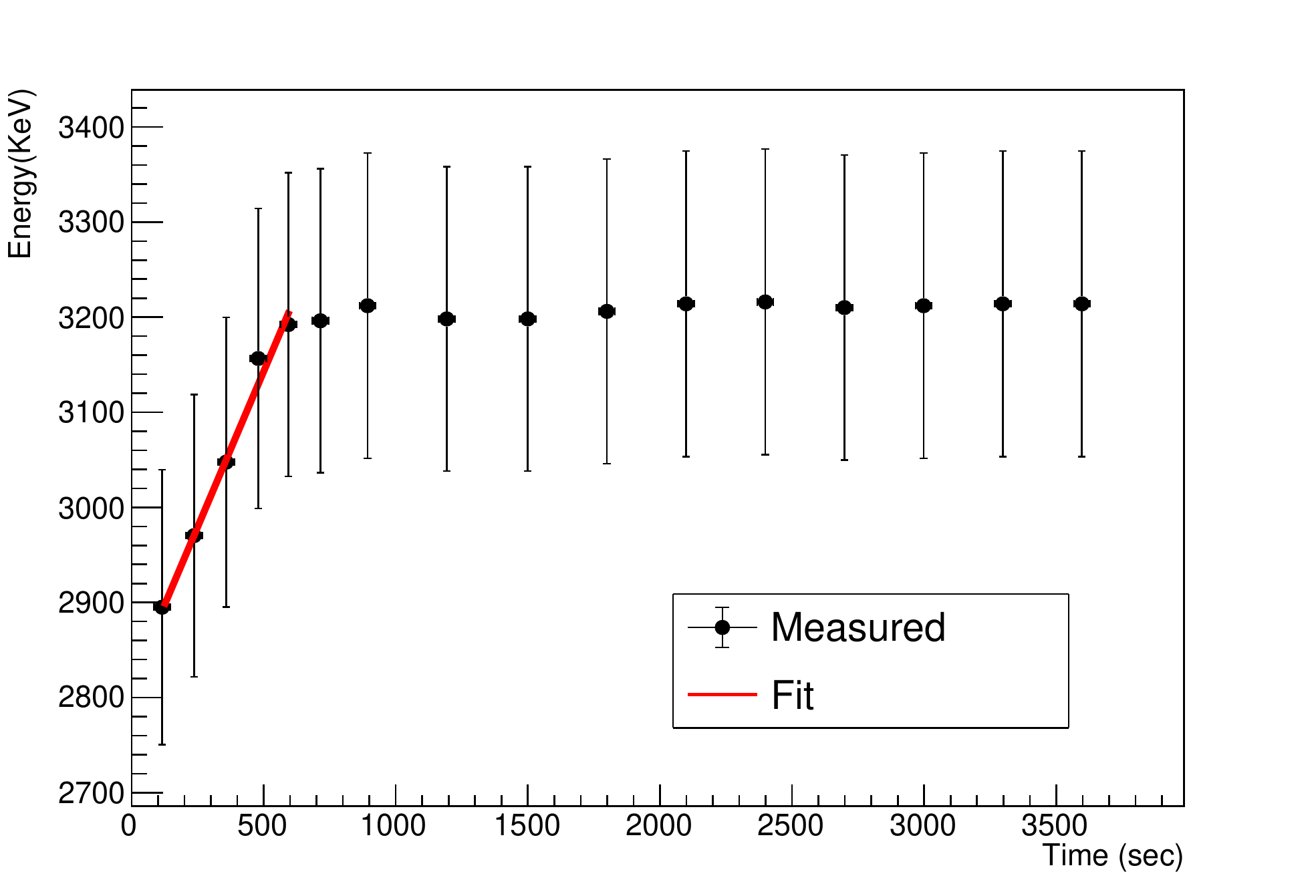}
\caption{The plot depicts the mean energy deposition ($E_{dep}$) as a function of time ($t$) for detector RL in Mode 2. As an illustration, the mean energy deposition ($E_{dep}$) and time ($t$) recorded for a bias voltage of $900$ volts are plotted for detector RL when operated in Mode 2. The error in $E_{dep}$ is attributed to the energy deposition determination using a Gaussian fit, while the error in $t$ is primarily due to the recorded time determination. A linear fit ($E_{dep} = p_0 \times t + p_1$) was applied to the part of the plot where the emission of charge carriers is greater than the trapping of charge carriers. The slope ($p_0$) of the fit was determined to be $654.17\pm2.76$ (eV/s), and the intercept ($p_1$) was found to be $2821.45\pm10.38$. It should be noted that the slope represents the emission rate of charge carriers ($e_n$) in equation~\ref{eq3}.}
\label{fig7}
\end{figure}

We conducted a measurement of energy deposition from $\alpha$-particles at 5.2 K over a period of 60 minutes for a given bias voltage to determine the charge emission rate mentioned in equation~\ref{eq3}. During this period, we captured the histogram of energy deposition every two to three minutes, and the mean energy deposition was obtained from the $\alpha$-peak. Figure~\ref{fig7} shows an example of this measurement for a bias voltage of 900 volts.

As depicted in Figure~\ref{fig7}, the application of bias voltage to the detector leads to a linear increase in the charge emission rate for the initial few minutes. This is because, in the initial minutes, de-trapping through impact ionization of dipole states or cluster dipole states outpaces the trapping of charge carriers at a given voltage. However, once the trapping and de-trapping reach a dynamic equilibrium, the energy deposition becomes constant. The slope of the plot's section, where the emission of charge carriers dominates, gives the charge-energy emission rate per unit of time, represented as $e_n$ in equation~\ref{eq3}. To convert the emission rate ($e_n$) into number of charge carriers, we divide $e_n$ by the binding energy of dipole states or cluster dipole states $(E_b)$. These emission rates in terms of the number of charge carriers per second are then used in equation~\ref{eq3} to determine the binding energy for respective dipole states or cluster dipole states. The calculated binding energies are presented in Table~\ref{tab:my_label1}. 

\begin{table*}[htp!!!]
\begin{adjustbox}{width=\textwidth,center=\textwidth}

    \begin{tabular}{|c|c|c|c|c|c|c|c|}
    \hline
    &&\multicolumn{3}{|c|}{Mode 1}&\multicolumn{3}{|c|}{Mode 2}\\
    \hline
        Bias voltage(V)& Electric field(V/cm)& Slope (eV/s)&Binding Energy(meV)&Trapping cross-section($cm^2$)&Slope(eV/s)& Binding Energy(meV)& Trapping cross  section($cm^2$)\\
        \hline
         
         -200&$186.92\pm4.02$&$116\pm5.80$&$8.105\pm0.405$&$(1.492\pm0.074)\times10^{-10}$&$1160\pm58$&$5.807\pm0.290$&$(4.62\pm0.231)\times10^{-11}$\\
         \hline
         -300&$280.37\pm4.01$&$89\pm4.45$&$7.916\pm0.395$&$(1.18\pm0.059)\times10^{-10}$&$1350\pm67.5$&$5.658\pm0.283$&$(3.59\pm0.179)\times10^{-11}$\\
         \hline
         -600&$560.74\pm3.98$&$119\pm5.95$&$7.239\pm0.362$&$(9.67\pm 0.484)\times10^{-11}$&$1820\pm91$&$5.429\pm0.271$&$(3.37\pm 0.168)\times10^{-11}$\\
         \hline
         -900&$841.12\pm4.03$&$164\pm8.2$&$6.898\pm0.345$&$(6.32\pm 0.316)\times10^{-11}$&$654\pm32.7$&$5.30\pm0.265$&$(2.32\pm0.116)\times10^{-11}$\\
         \hline
         -1100&$1028.04\pm4.01$&$-$& $-$&$-$&$871\pm$43.55&$5.237\pm0.262$&$(1.08\pm0.054)\times 10^{-11}$\\
         \hline
         -1200&$1121.50\pm4.03$&$271\pm13.55$& $6.553\pm0.328$&$(2.08\pm0.104)\times 10^{-11}$&$-$&$-$&$-$\\
         \hline
         
    \end{tabular}
    \end{adjustbox}
    \caption{The binding energy and trapping cross-section of RL at 5.2 K for Mode 1 and Mode 2. The errors, associated with
each value, are either the result of measurement errors or the error calculated from the equations used in the paper. Note that we could not get the data at  -1200 V for Mode 2 due to higher leakage current.}
    \label{tab:my_label1}   
\end{table*}

\begin{figure}[htp] 
\centering
\includegraphics[width=0.45\textwidth]{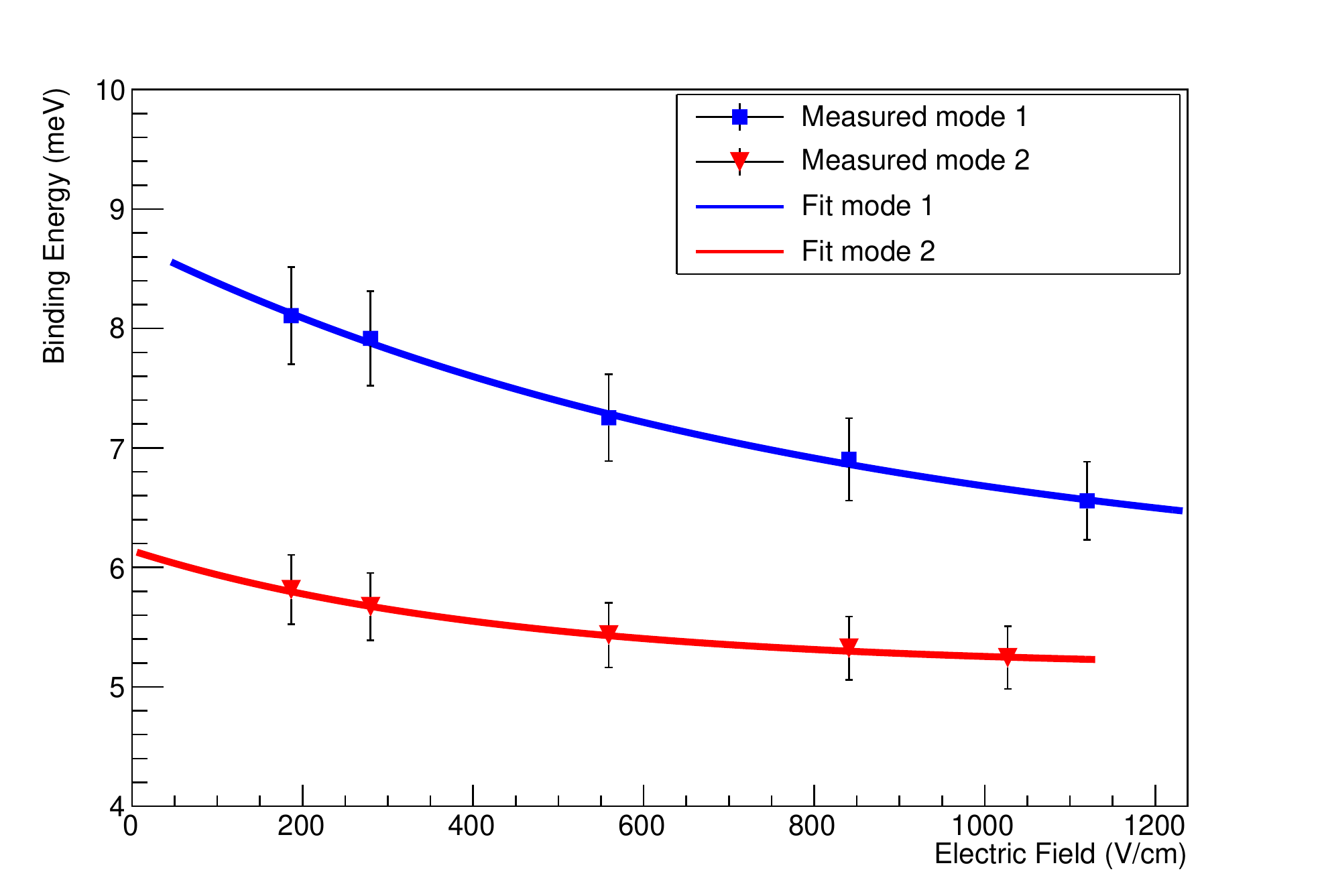}
\caption{The binding energies of the dipole states and the cluster dipole states have been determined as a function of the
applied electric field under two different operational modes,
Mode 1 and Mode 2. The error in the binding energy measurement was calculated, while the error in the electric field measurement was dominated by the precision of the applied bias voltage. To analyze the data, a fit model was used, specifically, $E_{bind}=(p_0)+(p_1)\times exp{-(p_2) \times (E)}$, which resulted in the following fitted parameters: For Mode 1, $p_0$ was found to be $5.845\pm 0.254$, $p_1$ was $2.871\pm 0.184$, and $p_2$ was $0.00123\pm 0.00003$. For Mode 2, $p_0$, $p_1$, and $p_2$ were $5.154\pm0.802$, $0.985\pm 0.063$, and $0.00229\pm0.000068$ respectively.}
\label{fig8}
\end{figure}

The binding energy of cluster dipole states is measured by the detector in Mode 2, while the binding energy of dipole states is measured by the detector in Mode 1. To measure the binding energy of cluster dipole states, we operate the detector in Mode 2, whereas for the binding energy of dipole states, we operate in Mode 1. Furthermore, the binding energy values obtained at different bias voltages demonstrate a correlation with the electric field. We have depicted this relationship in Figure~\ref{fig8}, where the binding energies are plotted as a function of the electric field at a temperature of 5.2 K.

The dipole states ($A^{0^{*}}$) in Mode 1 exhibit binding energies ranging from $6.553$ meV to $8.105$ meV, depending on the electric field. The average binding energy, computed as the sum of $p0$ and $p1$, is $8.716\pm 0.435$ meV when the electric field is zero. In contrast, the cluster dipole states ($A^{+^{*}}$) in Mode 2 have binding energies that range from $5.237$ meV to $5.807$ meV, depending on the applied electric field. The average binding energy at zero field is $6.138\pm0.308$ meV. Interestingly, the $A^{0^{*}}$ states have higher binding energy at zero fields than the $A^{+^{*}}$ states. Notably, both $A^{0^{*}}$ and $A^{+^{*}}$ states exhibit lower binding energies at zero field compared to the ground state impurity atoms in a Ge detector, which typically range in the order of 10 meV.

The populated dipole states and cluster dipole states with lower binding energies can be utilized to design a dark matter experiment with an extremely low-energy threshold, triggered by the small energy deposition through low-mass dark matter particles interacting with Ge atoms. This small energy deposition is dissipated through the emission of phonons that propagate through the detector volume and interact with dipole states or cluster dipole states, generating electron-hole pairs that are drifted towards electrodes. Our previous publication~\cite{Mei2018} demonstrates that if the detector's internal charge amplification of a factor of 100 to 1000 can be achieved, then the signal can be observed. It is important to note that these dipole states and cluster dipole states, which have lower binding energies, may also contribute to device noise, depending on the detector's operating temperature environment. At extremely low temperatures, such as 5.2 K, the probability of thermal excitation from these dipole states and cluster dipole states is very low due to the small thermal energy (0.45 meV) compared to the binding energies of these states (6-8 meV). Furthermore, the device noise is distributed across a broader range of energies, and their nature can be discerned and separated using techniques like low-pass or high-pass filters during calibration. By exciting the dipole and cluster dipole states, low-mass dark matter-induced events with lower energies can be statistically identified, depending on the event rate.

\section{Conclusion}\label{sec5}
Our investigation of binding energies and trapping cross sections in a p-type Ge detector at low temperatures has yielded important insights. Our measurements show that the binding energy of dipole states is $8.716\pm0.435$ meV, while the binding energy of cluster dipoles is $6.138\pm0.308$ meV, which is lower than the typical binding energy of ground state impurities in Ge. These binding energies are thermally stable at 5.2 K, and applying an electric field causes increased de-trapping via impact ionization for cluster dipoles due to their smaller binding energy compared to dipole states.

The trapping cross-section, ranging from $1.08\times 10^{-11} cm^2$ to $1.492\times 10^{-10} cm^2$, is primarily influenced by the electric field, with increasing electric fields leading to decreased binding energies and trapping cross-sections. These low binding energies suggest the possibility of developing a low-threshold detector for low-mass dark matter searches using appropriately doped impurities in Ge.

Overall, our findings provide valuable insights into the behavior of impurities in Ge detectors, which could inform the development of new detectors for dark matter searches and other applications.
\section{Aknowledgement}\label{sec6}
The authors are grateful to Mark Amman for providing instructions on how to construct planar detectors. We also acknowledge the contribution of a test cryostat from the Nuclear Science Division of the Lawrence Berkeley National Laboratory. This research was funded in part by NSF OISE 1743790, DE-SC0004768, and a South Dakota governor's research center.
\bibliography{mathbar.bib}


\end{document}